# Comment on "Partition Function of Anyon Gas"

G. Amelino-Camelia

Department of Physics
Boston University
590 Commonwealth Avenue
Boston, Massachusetts 02215 U.S.A.

and

Long Hua

Center for Theoretical Physics
Laboratory for Nuclear Science
and Department of Physics
Massachusetts Institute of Technology
Cambridge, Massachusetts 02139 U.S.A.
and
Department of Physics
New York University
New York, NY 10003 U.S.A.

In a recent Letter[1] Cho and Rim calculated the partition function of N anyons in an external harmonic well and/or in a constant external magnetic field. The calculation is based on the conjecture that all the energy eigenvalues depend linearly on the statistical parameter $\delta$ and that they belong to one of the "classes of eigenvalues" $E^I$ and $E^{II}$ given in eq.(5) of ref.[1]. This conjecture has been recently proven to be incorrect by several authors using different arguments.

In perturbative studies of 3 anyons in an external harmonic well, it was found that there are energy eigenvalues with nonlinear dependence on $\delta$. In ref.[2] using "fermionic end" perturbation theory (assuming $\delta \simeq 1$) it is shown analytically that there is an energy eigenvalue given to order $(1-\delta)^2$ by (we set $\hbar = c = 1$)

$$E^{ferm.} = \omega\{5 + \frac{9}{2}\ln(\frac{4}{3})(1-\delta)^2 + O((1-\delta)^3)\} \ . \tag{1}$$

In ref.[3] using "improved bosonic end" perturbation theory (assuming $\delta \simeq 0$) two more energy eigenvalues have been calculated analytically to order $\delta^2$

$$E^{bos.1} = \omega\{5 + \frac{9}{2}\ln(\frac{4}{3})\delta^2 + O(\delta^3)\}$$

$$E^{bos.2} = \omega\{6 - \frac{3}{2}\delta + \frac{9}{8}(3\ln(\frac{4}{3}) - 1)\delta^2 + O(\delta^3)\} \ . \tag{2}$$

Further indications that not all eigenvalues belong to the classes $E^I$ and $E^{II}$ have been given by numerical calculations of the low-lying energy levels for systems of three[4] and four[5] anyons in external harmonic well and constant magnetic field (these results are also mentioned in ref.[1]), and by the study[6] of the implications of the "modified conformal symmetry" which exists when an external magnetic field is applied to a system of anyons.

We also note that in ref.[7], where these classes are also discussed, it is suggested that $E^I$ and $E^{II}$ do not include all the energy eigenvalues.

In their Reply[8] Cho and Rim conjecture that the energies obtained in ref.[2-5] correspond to eigenfunctions which do not verify the hard core boundary condition. However, there seems to be no physical reason to impose this boundary condition to anyonic wave functions[9] and, anyway, both in "fermionic end" and in "improved bosonic end" perturbation theory, the 0-th order eigenfunctions do verify the hard core boundary condition, so that at any given perturbative order the anyonic wave functions also verify it. (The necessity of improving the 0-th order bosonic end eigenfunctions by imposing an hard core is discussed in ref.[2,3].)

Based on the results of ref.[2-7] it is easy to understand that the partition function obtained in ref.[1] is neither exact nor an approximation to the true partition function. In fact, among the energy eigenvalues which are ignored in the calculation there are some low-lying eigenvalues, which cannot be neglected even at low temperatures.

Our final comment concerns the equation of state obtained in ref.[1]. At present we cannot calculate the exact equation of state because the exact expressions for the eigenvalues which do not belong to $E^I$ and $E^{II}$ are not available. However, the results in ref.[2-7] indicate that, varying $\delta$ between 0 and 1, the complete set of anyonic energy eigenvalues interpolates continuously between the bosonic and the fermionic complete sets of eigenvalues; this would imply that the equation of state for a gas of anyons with $\delta \simeq 0$ ($\delta \simeq 1$) should be very similar to the one of a gas of bosons (fermions). On the contrary, in ref.[1] it is claimed that, for large N, a gas of anyons exhibits quantum behavior up to temperatures much higher than in the bosonic or fermionic case.

We thank C. Chou, R. Jackiw, and S.Y.-Pi for useful discussions. This work is supported in part by funds provided by the U.S. Department of Energy (D.O.E.) under contract #DE-AC02-89ER40509. G. A.-C. is supported partly by funds provided by the "Fondazioni Angelo Della Riccia", Firenze, Italy. L. H. would like to thank New York University for the Dean's Dissertation Fellowship and the Center for Theoretical Physics at MIT for the hospitality during this work.